# The Influence of Absorption and Gain on Photonic Density of States

A.H. Gevorgyan[1,2], K. B. Oganesyan[3,4] , N.Sh. Izmailian[3] , E.A. Ayryan[4],
Z. Mansurov[5], B. Lesbayev[5], G. Lavrelashvili[6] , M. Hnatic[7,8] ,
Yu.V. Rostovtsev[9] , G. Kurizki[10] , M.O. Scully[11]

[1]Yerevan State University, 1 Al. Manookian St., 025, Yerevan,Armenia
[2]Ins. of Applied Problems in Physics, 26, Hr. Nersessian, 0014, Yerevan, Armenia
[3]Alikhanyan National Science Lab, Yerevan Physics Institute, Alikhanyan Br.2, 036, Yerevan, Armenia, bsk@.yerphi.am
[4]Laboratory of Information Technologies, JINR, Dubna, Russia
[5]Institute of Combustion Problems, Almaty, Kazakhstan
[6]A.Razmadze Mathematical Institute, Tbilisi, Georgia
[7] Faculty of Sciences, P. J. Safarik University, Kosice, Slovakia
[8]Institute of Experimental Physics SAS, Kosice, Slovakia
[9]University of North Texas, Denton, TX, USA
[10]Weizmann Institute of Science, Rehotot, Israel
[11]Texas A&M University, College Station, Texas, USA

## Abstract

Using the exact analytic expressions for the reflection and transmission matrices for the finite thickness cholesteric liquid crystal (CLC) layer we calculated its photonic density of states (PDS) of the eigen polarizations (EPs). We investigated the influence of absorption and gain, as well as the CLC cell thickness and CLC local dielectric anisotropy on PDS. We presented the full picture of distribution of total field in the CLC layer and outside it for two EPs. The possibility of connections between the PDS and the density of the light energy accumulated in the medium is investigated and it was shown that these characteristics have analogous spectra and, besides, the influences of the problem parameters on these characteristics also are analogous. We showed that there exists a critical value of the parameter characterizing the gain beyond which the lasing mode is quenched and the feedback vanishes. We showed that in the presence of gain there exists a critical value of numbers of pitches in CLC layer beyond which the lasing mode again is quenched and the feedback vanishes, too. It is shown that the subject system can work as a low threshold laser or a multi-position trigger.

## 1. Introduction

Recently there has been great interest in photonics and it has been energetically developing. Photonics is the science and technology of generating and controlling photons, particularly in the ultra-violet, visible and infra-red light spectrum. Photonic crystals (PCs) and metamaterials, with the potential of allowing comparable control over light, are the subject of intense interest today, too. They have been widely used as laser cavities. Dowling et al. predicted [1] lasing at the band edges of the photonic band gap (PBG) materials based on the argument that light slows down near the band edge and thus the spontaneous emission can be enhanced. The Dowling mechanism of lasing applies to only infinite or very long systems. In [2], it was discovered that cholesteric liquid crystals (CLCs) are PBG materials. Due to the spontaneously-formed helical structure these CLCs act as one-dimensional PCs that exhibit many interesting tunable optical properties. In the same work, it was experimentally discovered the lasing in finite systems, such as CLCs, where the lasing occurs in specific modes, which have very different lifetimes depending on how close they are to the band edge, and this gives selectivity for lasing in long-lived modes. It should be noted that it is not the light with a low velocity which is involved, but essentially the standing waves associated with the standing resonances. We should also note that to explain the absorption peculiarities in CLCs the peculiarities of the total standing waves aroused in the CLC layer and the light energy distribution inside the system, as well as the spectra peculiarities of the group velocity in PCs were investigated in [3-5]. It was shown in those works that a standing wave having a modulation peculiarity along the axis, $z$ ($z$ is directed into the depth of the system) is aroused in the system. The wave in the PBG is evanescent and it exponentially decreases along $z$, leading to suppressing of the absorption. An anomalously strong absorption takes place near the PBG borders, on the reflection minima. Then it was shown that at those wavelengths both a decrement of the group velocity and light energy accumulation in the medium take place. The converse phenomena take place outside the PBG at the reflection maxima. Below we investigate the possibility of existence of relation between PDS and density of light energy.

Let us note that tuning light emission by the use of the engineered nanophotonic structures (PCs and metamaterials) is an active area of research due to its potential applications in single photon generation [6], miniature lasers [7], light emitting diodes [8], solar energy harvesting [9], and etc. This can be achieved owing to the possibility of tuning the photonic density of states and hence the spontaneous emission by engineering the nanophotonic structures [10] or by external fields. The CLCs are one of the most interesting types of 1D PC, because of easy tenability of their parameters. The performance of light emitting liquid crystal devices may be significantly improved by using more advanced optical structures. In order to develop new optical architectures a fundamental understanding of the emission process and accurate numerical design tools are necessary.

The lasing in the CLCs and PCs is being intensely investigated (see, for instance, [11, 12] and the wide literature cited there). The possibilities of decreasing the lasing threshold in the CLC and multilayer systems with CLC layers have been studied theoretically [13-15] and experimentally [16]. In the following two papers [17, 18] it was reported the observation of a continuous wave lasing in dye doped CLCs and the effect of loss and gain of the photonic density of states (PDS) was calculated there.

In this paper we discussed the influence of: absorption; gain; layer thickness; local dielectric anisotropy and etc on the PDS. The possibility of a connection between the PDS and the density of the light energy accumulated in the medium was investigated. All these were done on the base of the exact analytical expressions of the reflection and transmission matrices, which were obtained by solving the boundary problem for the CLC finite planar layer [19, 20].

## 2. Method of analysis

To investigate the PDS peculiarities of a single CLC layer we used the exact analytical expressions of the reflection and transmission matrices for the CLC finite planar layer obtained in [19, 20]. We investigate the peculiarities of the PDS for the EPs.

The EPs are the two polarizations of the incident light, which do not change when light transmits through the system [21-23]. The EPs and eigen values (EVs, that is, the amplitude transmission and reflection coefficients for the incident light with the EPs) deliver much information about the peculiarities of light interaction with the system. Therefore, the calculation of EPs and EVs of every optical device is important. It follows from the definition of the EPs that they must be connected with the polarizations of the excited internal waves (the eigen modes) aroused in the medium. (In the majority of cases they coincide with the polarizations of the eigen modes). Naturally, there are certain differences in the general case: there exist only two EPs; meanwhile, the number of eigen modes can be more than two, and the polarizations of these modes can differ from each other (for the non-reciprocity media, for instance).

The EPs automatically take the influence of the dielectric borders into account. As it is known (in particular, for the normal incidence) the EPs of either CLC or gyrotropic media practically coincide with the orthogonal circular polarizations; meanwhile, they coincide with the orthogonal linear polarizations for the non-gyrotropic media. It follows from the above-said that the investigation of the EPs peculiarities is especially important for the case of inhomogeneous media for which the exact solution of the problem, in general, is unknown.

Denoting the ratio of the complex field components by $\chi_i = E_i^- / E_i^+$ for the incident wave at the entrance of the system, and the same ratio at the exit by $\chi_t = E_t^- / E_t^+$, and taking into account that:

$$\begin{bmatrix} E_t^+ \\ E_t^- \end{bmatrix} = \begin{bmatrix} T_{11} & T_{12} \\ T_{21} & T_{22} \end{bmatrix} \begin{bmatrix} E_i^+ \\ E_i^- \end{bmatrix},$$

we find that:

$$\chi_t = (T_{22}\chi_i + T_{21})/(T_{12}\chi_i + T_{11}), \tag{1}$$

where $T_{ij}$ are the elements of the total system transmission matrix.

The function, $\chi_t = f(\chi_i)$, is called *polarization transfer function* [23] and it bears information about polarization ellipse transformation when light is transmitted through the system. Every optical system has two EPs, which are obtained through substitution $\chi_t$ from equation (1) into the equation $\chi_i = \chi_t$. Thus we get:

$$\chi_{1,2} = \frac{T_{22} - T_{11} \pm \sqrt{(T_{22} - T_{11})^2 + 4T_{12}T_{21}}}{2T_{12}}. \tag{2}$$

The ellipticities, $e_{1,2}$, and the azimuths, $\psi_{1,2}$, of the EPs are expressed by $\chi_{1,2}$ through the following formulae:

$$\psi_{1,2} = -\frac{1}{2}\arg(\chi_{1,2}), \qquad e_{1,2} = arctg\left(\frac{|\chi_{1,2}| - 1}{|\chi_{1,2}| + 1}\right). \tag{3}$$

## 3. Results and discussion

Now we pass to the results. We investigate the PDS spectra peculiarities for the two EPs. The ordinary and extraordinary refractive indices of the CLC layers are taken to be $n_o = \sqrt{\varepsilon_2} = 1.4639$ and $n_e = \sqrt{\varepsilon_1} = 1.5133$ ($\varepsilon_1$, $\varepsilon_2$ are the principal values of the CLC local dielectric tensor), the CLC layer helix is right handed and its pitch is: $p$ = 420 nm. These are the parameters of the ***CLC cholesteryl-nonanoate–cholesteryl chloride–cholesteryl acetate*** (20 : 15 : 6) composition, at the temperature $t$ = 25°C. So, the light normally incident onto a single CLC layer – with the right circular polarization (RCP) has a PBG (which is in the range of λ = 614.8 ÷ 635.6 nm) and the light with the left circular polarization (LCP) does not have.

We consider the peculiarities of the PDS for the above-said CLC cell. The divergence in the PDS at band edges in the CLC and at the defect modes in the CLC with defects in its structure was shown theoretically and experimentally in [2, 14, 15, 17, 24]. Some new and important peculiarities of the PDS of the CLC layer as well as of the CLC-NLC structure were obtained in [17, 25-27]. The investigation of the PDS peculiarities is important because of the following. For laser emission it has been shown that, for instance, when analyzing the case of the Fabry-Perot resonator, that the threshold gain $g_{th}$ can be related directly to the PDS $\rho$ as [27]:

$$g_{th} \propto \frac{n}{d\rho}, \tag{4}$$

where $n$ is the refractive index inside the resonator of the length, $d$, and $\rho$ is the maximum PDS. Furthermore, according to the space-independent rate equations, the slope efficiency of lasers can be shown to be inversely proportional to the threshold energy and, therefore, directly proportional to $\rho$ [28].

The PDS is converse to the group velocity and is defined by the expression [28, 29]:

$$\rho_i(\omega) \equiv \frac{dk_i}{d\omega} = \frac{1}{d} \frac{\frac{du_i}{d\omega}v_i - \frac{dv_i}{d\omega}u_i}{u_i^2 + v_i^2}, \; i = 1,2 \tag{5}$$

where $d$ is the CLC layer thickness, and $u_i$ and $v_i$ are the real and imaginary parts of the transmission coefficients: $T_i(\omega) = u_i(\omega) + iv_i(\omega)$ are the transmission coefficients for the incident light with the first and second EPs, respectively. The exact expressions for the reflection coefficient, $R_i(\omega)$, and that for the transmission, $T_i(\omega)$, for the normal light incidence on the CLC planar layer were obtained in [19, 20]. These expressions are very complicated for the general case, but for the following cases: $\varepsilon_s = \varepsilon_m$ and $d = ip$ ($\varepsilon_s$ the dielectric permittivity of the medium bordering the CLC layer on its both sides, $\varepsilon_m = \frac{\varepsilon_1 + \varepsilon_2}{2}$, $\varepsilon_1$, $\varepsilon_2$ are the principal values of the CLC local dielectric tensor and $i$ is an integer) they are very simple. Taking the following form for the incident wave:

$$\vec{E}_i = E_{i1}\vec{e}_1 + E_{i2}\vec{e}_2 \tag{6}$$

where $\vec{e}_{1,2} = \begin{pmatrix} 1 \\ \beta_{1,2} \end{pmatrix}$ are the EP orts, one can have for the transmitted and reflected waves:

$$\vec{E}_t = V_1 E_{i1}\vec{e}_1 + V_2 E_{i2}\vec{e}_2, \quad \vec{E}_r = W_1 E_{i1}\vec{e}_{r1} + W_2 E_{i2}\vec{e}_{r2}, \tag{7}$$

($\vec{e}_{r1,2} = \begin{pmatrix} 1 \\ \beta_{r1,2} \end{pmatrix}$ are the EP orts of the reflected wave). According to [20], we obtain for the coefficients, $V_{1,2}$, $W_{1,2}$, $\beta_{1,2}$ and $\beta_{r1,2}$ (if $\varepsilon_s = \varepsilon_m$ and $d = ip$):

$$V_{1,2} = \frac{1}{a_{2,1}}, \; W_{1,2} = \frac{iu\delta s_{2,1}\beta_{1,2}}{a_{2,1}|\beta_{1,2}|}, \; \beta_{r1,2} = \frac{1}{\beta_{1,2}}, \quad \beta_{1,2} = \frac{2\chi \pm \gamma}{\delta}, \tag{8}$$

where $a_{1,2} = c_{1,2} \mp iul_{1,2}s_{1,2}$, $l_{1,2} = \gamma \pm 2$, $s_{1,2} = \frac{sin(k_{1,2}d)}{k_{1,2}d}$, $c_{1,2} = cos(k_{1,2}d)$, $k_{1,2} = \frac{2ub^{\pm}}{d}$,

$b^{\pm} = \sqrt{1+\chi^2 - \delta \pm \gamma}$, $\gamma = \sqrt{4\chi^2 + \delta^2}$, $\delta = \frac{\varepsilon_1 - \varepsilon_2}{\varepsilon_1 + \varepsilon_2}$, $u = \frac{\pi d \sqrt{\varepsilon_m}}{\lambda}$, $\chi = \frac{\lambda}{p\sqrt{\varepsilon_m}}$, λ is the wavelength in vacuum. Thus, we have, $T_i(\omega) = V_i(\omega)$. Taking (8) into account we have for the PDS of the EPs:

$$\rho_{1,2} = \frac{\sqrt{\varepsilon_m}\left\{\chi^2\delta(\delta+4)c_{1,2}s_{1,2} + l_{1,2}\left[\gamma(1-\delta) \pm (\delta^2 + 2\chi^2)\right]\right\}}{2\gamma b^{\pm 2}\left(c_{1,2}^2 + u^2 l_{1,2}^2 s_{1,2}^2\right)c}. \tag{9}$$

For the isotropic case, we have: $\rho_{iso} = \frac{n_s}{c}$, where $n_s$ is the refractive index of the CLC layer surrounding the medium and *c* is the speed of light in a vacuum. These expressions hold for the case if absorption or gain is absent as well as if the absorption or gain is present. Both loss and gain can be incorporated through introducing an imaginary part to the CLC dielectric permittivity tensor components.

Now we investigate the absorption (emission) peculiarities of the CLC layer. Let the CLC layer be doped with dye molecules. Then this system is an amplifier in the presence of a pumping wave, i.e. we are discussing a planar resonator with active elements. The presence of dye leads to the change of the local refraction indices of the system. In this case, the effective imaginary parts of the local refraction indices in the CLC ($n''_{1,2}$) are negative ($n_{1,2} = \sqrt{\varepsilon_{1,2}} = n'_{1,2} + in''_{1,2}$), and they all characterize the gain. In the case of absorption (when the imaginary parts of the local CLC indices, $n''_{1,2}$, are positive), the quantity, $A = 1-(R+T)$, characterizes the light energy absorbed by the system, while, in the case of amplification, the value of $|A|$ characterizes the radiation emitted by the system (we assume that the incident light intensity is unit, $I_0 = 1$).

We characterize the degree of arrangement of dipole transition momentums of the guest molecules by the order parameter, $S_d$, defining it through the average of cosϑ:

$$S_d = \frac{3}{2}\langle\cos\vartheta\rangle - \frac{1}{2}, \tag{10}$$

where ϑ is the angle between the local optical axis and the dipole transition momentum of the guest molecule. The possible maximum value of this order parameter is the unit, $S_d = 1$, and it corresponds to the ideal orientation of transition dipole momentums along the local optical axis; the isotropic orientation of transition dipole momentums corresponds to the value, $S_d = 0$, and the minimum value, $S_d = -0.5$, corresponds to the distribution of those momentums in the plane, which are perpendicular to the local optical axis. For more details, see [31-100].

It is also to be noted that, as it was shown in [30], if taking the Maxwellian boundary conditions for plane wave transmission through the finite thickness amplifying layer into account and using a dielectric permittivity with a negative imaginary part, when the field radiation is small (that is, if $|E|^2 << \frac{\hbar^2}{|\mu|^2 \tau_p \tau_n}$, where $\mu$ is the dipole momentum of transition of the quantum point; $\tau_p$ and $\tau_n$ are the characteristic relaxation times of polarization and inverse inhabitations, respectively) then Fresnel's approach based on the solution of the wave equation gives results coinciding with those numerically modelized according to Maxwell-Bloch's theory.

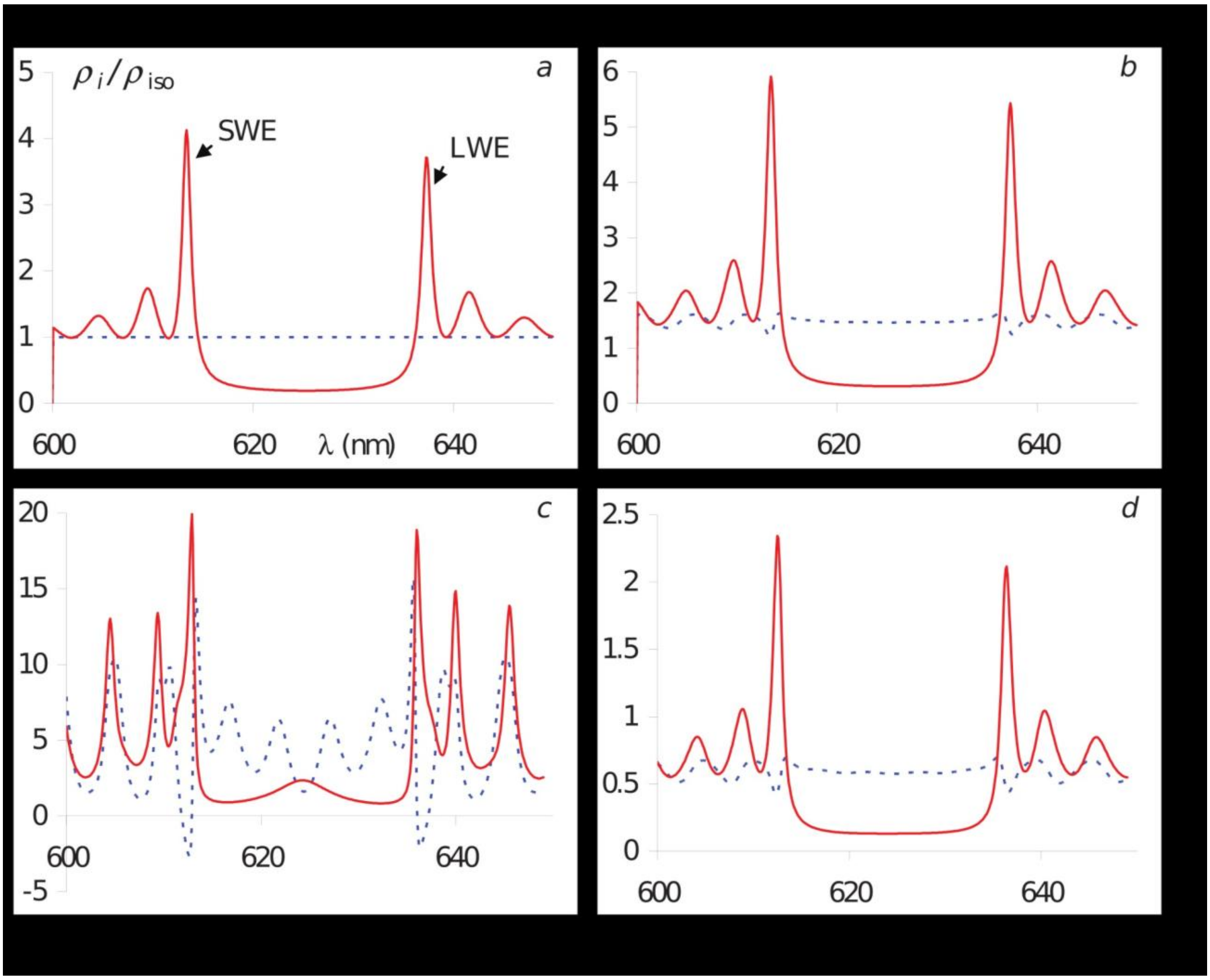

Fig. 1. The spectra of $\frac{\rho_i}{\rho_{iso}}$ (i.e. the normalized PDS) for different values of $\varepsilon_s$ (i.e. for the CLC layer surrounding media dielectric permittivity). *a*: $\varepsilon_s = \varepsilon_m$; *b*: $\varepsilon_s = 1$; *c*: $\varepsilon_s = 0.1$; *d*: $\varepsilon_s = 6.25$. The CLC layer thickness is: $d$ = 50$p$. The light incident on the CLC has a diffracting EP ($i$ = 2, the red solid lines) and a non-diffracting EP ($i$=1, the blue dashed lines).

In Fig. 1 the spectra of $\rho_i / \rho_{iso}$ for various values of $\varepsilon_s$ are presented. The red line is for the diffracting EP, and the blue line is for the non-diffracting EP. As it is seen from Fig. 1*a* and as it is well known, $\rho_i / \rho_{iso}$ for the diffracting EP has two principal maxima near the PBG borders (the short wavelength edge (SWE) maximum and the long wavelength edge (LWE) maximum). We can see from the plots, that the PDS of diffracting EP increases dramatically with decreasing of $\varepsilon_s$ (from the value, $\varepsilon_s = \varepsilon_m$), and vise verse, it decreases dramatically with increasing of $\varepsilon_s$ (again from the value, $\varepsilon_s = \varepsilon_m$). Additionally, the PDS of non-diffracting EP does not show any changes with λ at $\varepsilon_s = \varepsilon_m$, meanwhile, it carries some changes at $\varepsilon_s \neq \varepsilon_m$ and, at smaller values of $\varepsilon_s$, it oscillates with high amplitudes – particularly, near the PBG borders.

Now we pass to the investigation of the influence of absorption and gain on the PDS. In Fig. 2 the dependences of the maximum, $\rho_{2\max} / \rho_{iso}$, for the diffracting EP at the short (the blue solid line) and long (the red dashed line) wavelength edges on the absorption (*a, c, e*) and gain (*b, d, f*) are presented. The first row is for the case of the isotropic absorption and gain ($S_d = 0$); the second row corresponds to the case of the ideal orientation of transition dipole momentums along the local optical axis ($S_d = 1$) and, finally, the third row is for the case of the isotropic distribution of those momentums in the plane, which are perpendicular to the local optical axis ($S_d = -0.5$).

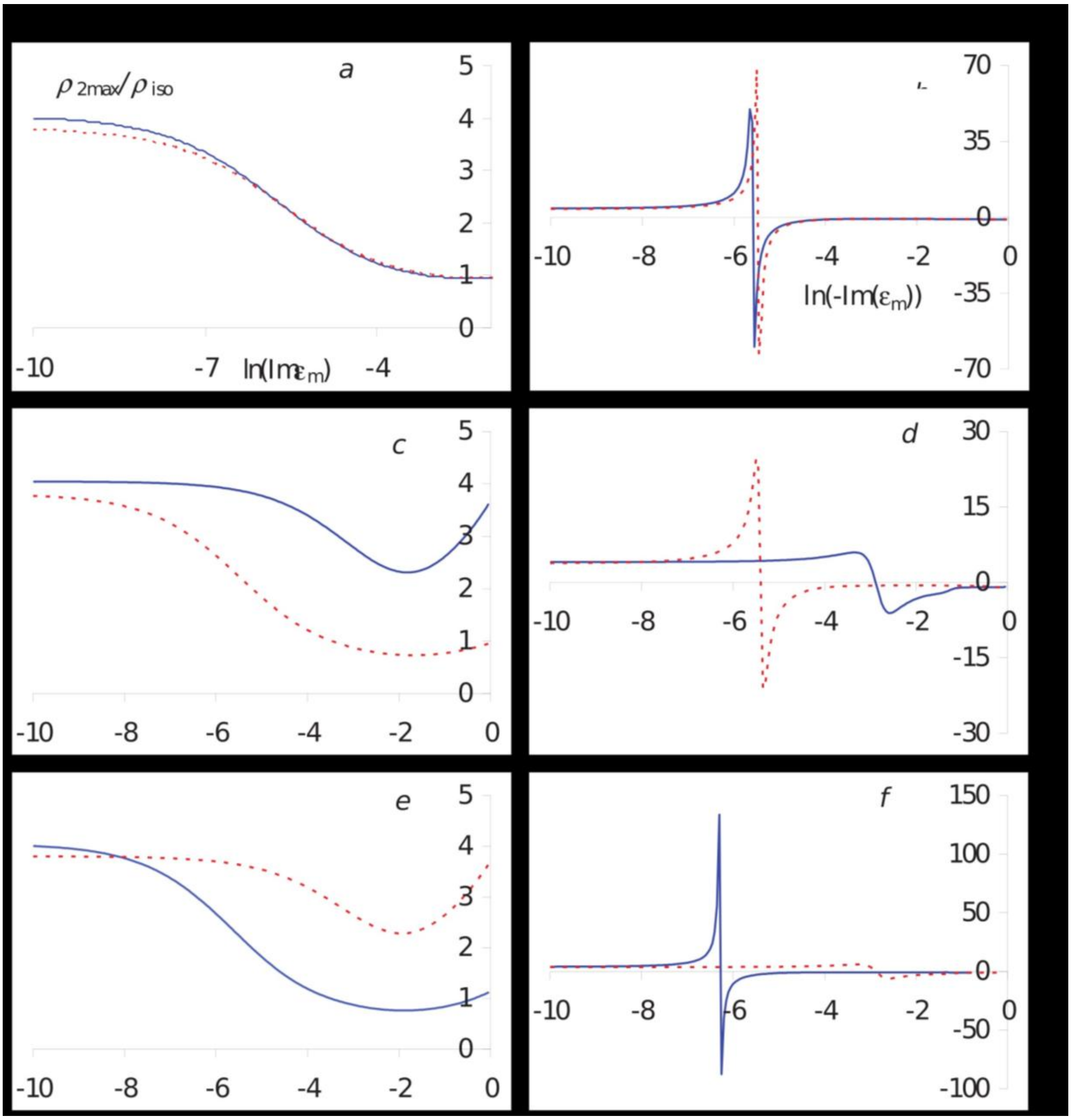

Fig. 2. The dependence of the maximum PDS ( $\rho_{2\max} / \rho_{iso}$ ) for the diffracting EP at the short (the blue solid line) and long (the red dashed line) wavelength edges on the absorption (*a, c, e*) and gain (*b, d, f*). The first row corresponds to the isotropic absorption and gain ($S_d = 0$), the second row corresponds to the case of the ideal orientation of the transition dipole momentums along the local optical axis ($S_d = 1$) and finally, the third row is for the case of the distribution of those momentums in the plane perpendicular to the local optical axis ($S_d = -0.5$). $d = 50p$, $\varepsilon_s = \varepsilon_m$ .

As it can be seen from the plot (Fig. 2*a*), the PDS decreases dramatically with increasing of the parameter, $x = \ln(\mathrm{Im}\,\varepsilon_m)$ , characterizing the absorption. It is a well known result (see, for instance, [17]) and it is called forth by an anomalously strong light absorption at the corresponding wavelengths, leading to a sharp decrement of the density of the light energy accumulated in the system (see the details below). The case of the anisotropic absorption is interesting, too. There are two important effects here.

(a) Small absorption. If an imaginary term is included only in the dielectric constant parallel to the local director, then the PDS in the SWE is practically unaffected when the loss is increased, while the PDS in the LWE is diminished. Meanwhile, if an imaginary term is included only in the dielectric constant perpendicular to the local director, then the reverse phenomenon takes place.

(b) Further increase of the loss. If an imaginary term is included only in the dielectric constant parallel to the local director, then the PDS in the short wavelength edge dramatically decreases at first and then it dramatically increases. For the higher values of the anisotropic absorption, the PDS increases at the LWE, too. If an imaginary term is included only in the dielectric constant perpendicular to the local director, then the reverse phenomenon takes place.

To find out the physical cause of the above mentioned effects one has to take the modulation peculiarities of the total wave aroused in the CLC layer into account (see Fig. 3; about the method of calculation of the total field aroused in the CLC layer and outside it, see below). As it is known and as it is presented in Fig. 3, in the case of normal light incidence and if it has the diffracting EP, the total standing wave field aroused in the medium is linearly polarized in the PBG and, moreover, the total field direction is fixed for a fixed coordinate along the axis of the helix. As the coordinate *z* (directed along the medium axis) changes, this direction rotates around *z* in such a way that the angle between the director at each point and this direction remains the same. If the light frequency changes, the angle between the director and total field also changes. Note that the jump change of the azimuth, $\psi$, nearby of the PBG centre is due to the fact that the function, arctg, is defined in the interval, ($-\pi/4 \div +\pi/4$). At the short-wavelength boundary of the PBG, the total field appears to be aligned along the direction corresponding to the minor principal value of the local permittivity tensor $\hat{\varepsilon}$. At the long-wavelength boundary of the PBG, the field is directed across the direction of the minor principal value. It is evident that introduction of the absorption, which we assume for definiteness to be small, at first, will not considerably affect the polarization characteristics of the eigen waves. For this reason, even in the presence of absorption, the field structure described above remains the same. It follows from this that, if an imaginary term is included only in the dielectric constant parallel to the local director (only in $\varepsilon_1$), then the smallest absorption will be observed at the short-wavelength boundary of the PBG and, therefore, the PDS in the SWE is practically unaffected while at the long-wavelength boundary an anomalously strong absorption is observed and, therefore, PDS is diminished here. And vice-versa, if an imaginary term is included only in the dielectric constant perpendicular to the local director (only in $\varepsilon_2$), then the smallest absorption is observed at the long-wavelength boundary of the PBG and, therefore, the PDS in the LWE is practically unaffected.

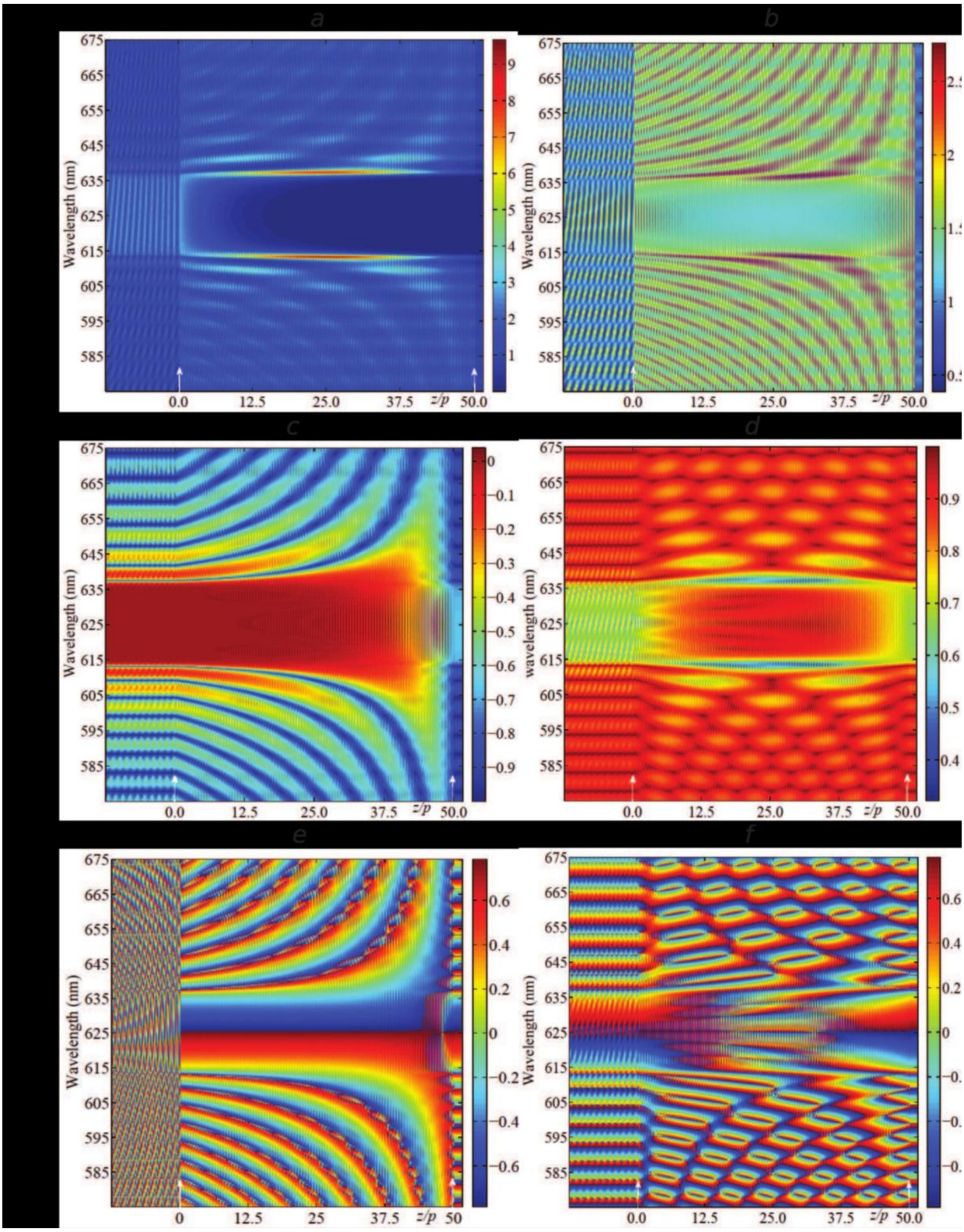

Fig.3. (Color online). The density plot of the $\left|\vec{E}^{0,1,2}\right|^2$ (the first row), the total wave ellipticity, $e$ (the second row) and the azimuth, $\psi$ (the third row) spectra as functions of $z/p$ for the diffracting (the left column) and non-diffracting (the left column) EPs at the absence of absorption and gain. $n_s = \sqrt{\varepsilon_m}$ ($n_s$ is the CLC layer surrounding media refractive index). The CLC borders are denoted by the white arrows in the figure.

Now, let us go back to Fig. 2. If the absorption is again increased, then the above presented description changes. Now one must take the following into account. When the absorption is anisotropic, a periodic change of absorption is added to the periodic change of refraction. Therefore, in this case (that is, in the case, if the absorption is anisotropic), when the parameter *x* increases, then both the absorption and the efficiency of diffraction on the periodic spatial change of absorption are increased, too. And the increment of the diffraction efficiency – as it is known – leads to an increment of the maximum PDS (see, Fig. 2 *c,e*). In [17], the increment of the maximum PDS was shown when the refraction anisotropy increases; below we investigate, in detail, the influence of the refraction anisotropy on the PDS, too.

When the gain (that is, the parameter, $x' = \ln(-\mathrm{Im}\varepsilon_m)$ increases, the maximum PDS is increased too. The further increment of the gain leads to a resonance-like change of the maximum PDS and then the PDS is diminished. There exists a critical value of $x'$ beyond which the lasing mode is quenched and the feedback vanishes. Note that the critical values of $x'$ for the SWE and LWE are different. The existence of these critical values of the gain indicates on the possibility that the subject system can work as a trigger. Tuning the pumping intensity one can (choosing a suitable dye molecule density) pass from the laser generation regime to the regime of quenching.

To complete this account we present in Fig. 4 the evolution of the PDS spectra when the absorption and gain increase. Let us note here one important and very interesting result. When the gain increases, the maximum PDS goes away from the PBG border, and this does not take place continuously, but in discrete paces (Fig. 4*b,d*). Note that when going away from the PBG borders, the critical value of $x'$ (beyond which the lasing mode is quenched) is increased. To realize the obtained results one has to take the following into account. As it is well known, for the CLC layer of a finite thickness the reflection coefficient equals to the unit (in the PBG) and it does not smoothly decrease outside the PBG if the wavelength is far from the PBG. Instead, it oscillates. These oscillations are observed over and over again experimentally, too. These oscillations are due to the finiteness of the CLC layer thickness and, therefore, are the consequence of light diffraction inside the finite volume and they are not connected with the reflections from the dielectric borders. The presence of the dielectric borders (i.e. the difference of $n_s$ from $n_m = \sqrt{\varepsilon_m}$ ) leads to an additional modulation of the oscillations outside PBG.

Thus, there are two mechanisms of diffraction when light interacts with a finite PC layer. One is due to light diffraction on a periodic PC structure and it manifests itself, arousing regions of diffraction reflection (or PBG), and the other one is due to the light diffraction in the finite volume (owing to finiteness of the PC layer volume) and this one manifests itself nearby the PBG, arousing diffraction oscillations like the pendulum beatings (which are well known in X-ray diffraction). One must clearly distinguish these oscillations from the interferential oscillations from the dielectric borders, which are also observed when light transmits through a homogeneous medium layer. They have different periods for both the frequency and thickness. Oscillations of diffraction are also observed if the dielectric borders have the minimum influence, i.e. if $n_m = n_s$. Meanwhile, no interferential oscillation from the dielectric borders is observed in this case. The minima of the reflection approximately are defined from the condition:

$$Kd_1 = m\lambda, m = 0,1,2,..., \tag{11}$$

where $K = \sqrt{\frac{\omega^2}{c^2}\frac{\varepsilon_1+\varepsilon_2}{2} + a^2 \pm \sqrt{\left(\frac{\omega^2}{c^2}\frac{\varepsilon_1-\varepsilon_2}{2}\right)^2 + 4a^2\frac{\omega^2}{c^2}\varepsilon_m}}$, $a = 2\pi/p$.

And the PDS has maximum values just for the modes defined by condition (11). This is the cause of the above peculiarities.

Comparison of the results presented in Fig. 4 with those in Fig. 2 and Fig. 3 shows that the subject system can work as a multi-position trigger.

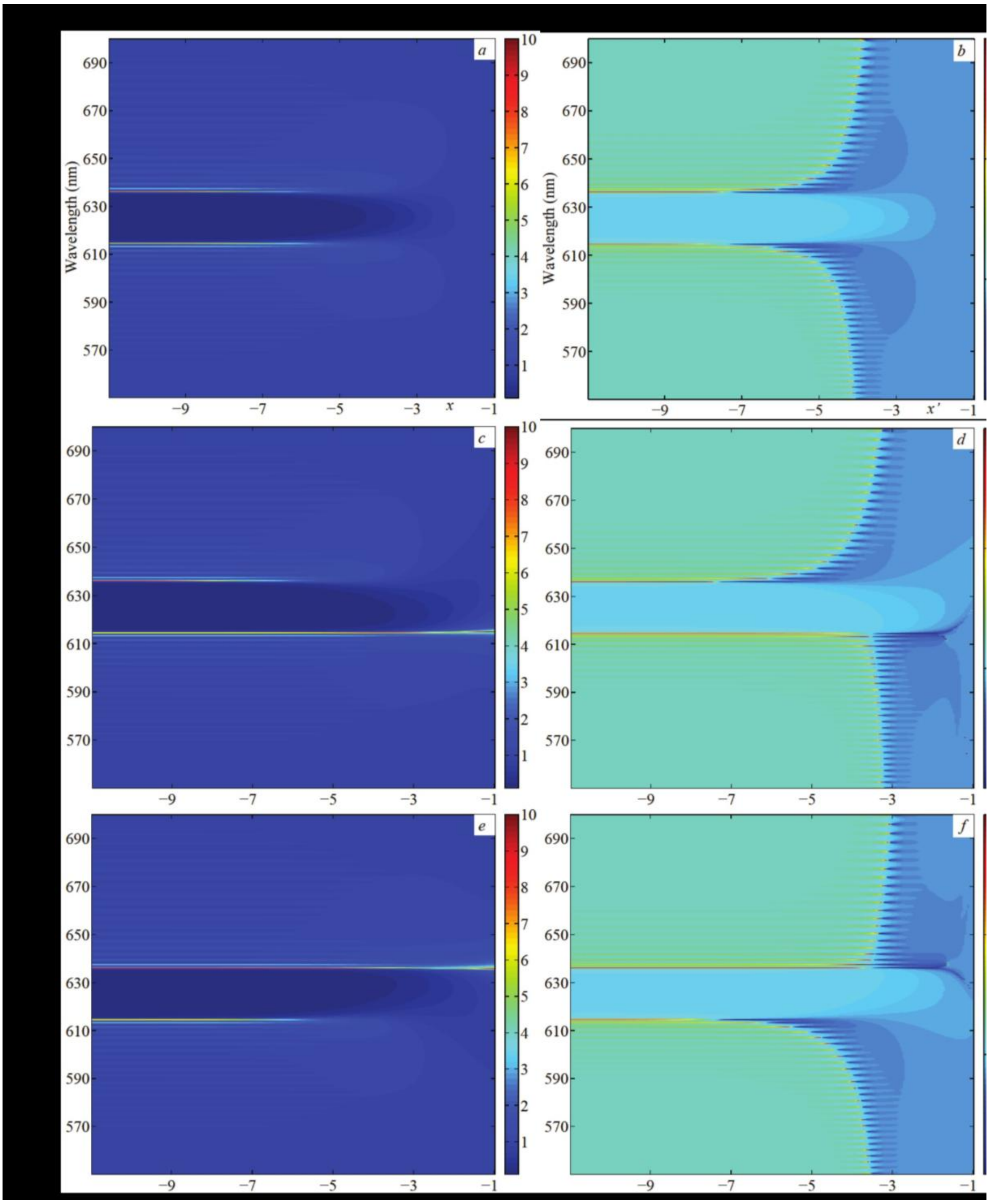

Fig. 4. (Color online). The density plot of the PDS spectra as a function of the parameters $x$ (the left column) and $x'$(the right column) characterizing the absorption ($x$) and gain ($x'$).The first row corresponds to the isotropic absorption and gain case ($S_d$ = 0), the second row corresponds to the case of the ideal orientation of transition dipole momentums along the local optical axis ($S_d$ =1) and the third row is for the case of the distribution of those momentums in the plane perpendicular to the local optical axis ($S_d$ = –0.5). The incident light has a diffracting EP. $d$ = 50$p$, $\varepsilon_s = \varepsilon_m$.

Now we pass to the further detailed investigation of the total wave distribution inside/outside the CLC layer, also investigating the modulation peculiarities along the axis, $z$ ($z$ is directed into the depth of the medium), as well as to the investigation of the polarization peculiarities of the total wave. We investigate the spectra peculiarities of the density of the light energy and this density relation with the PDS.

The total field in each medium (we discuss a CLC layer sandwiched between two isotropic half spaces) can be presented as follows:

$$\vec{E}^0 = \vec{E}_i + \vec{E}_r, \ \vec{E}^1 = \vec{E}^{in}, \ \vec{E}^2 = \vec{E}_t, \tag{12}$$

where the indices, 0, 1, and 2, denote the fields corresponding to the media on the left-hand side of the CLC layer, the layer, and the medium on the right-hand side of the layer, respectively; $\vec{E}_i$, $\vec{E}_r$ and $\vec{E}_t$ are the fields of incident, reflected and transmitted waves, respectively; and $\vec{E}^{in}$ is total field in the CLC layer. They are defined by the boundary conditions. The averaged light energy density in each medium is calculated as follows:

$$w^{0,1,2} = \frac{1}{4\pi V} \int n_{0,1,2}^2 \left| \vec{E}^{0,1,2} \right|^2 dV \, . \tag{13}$$

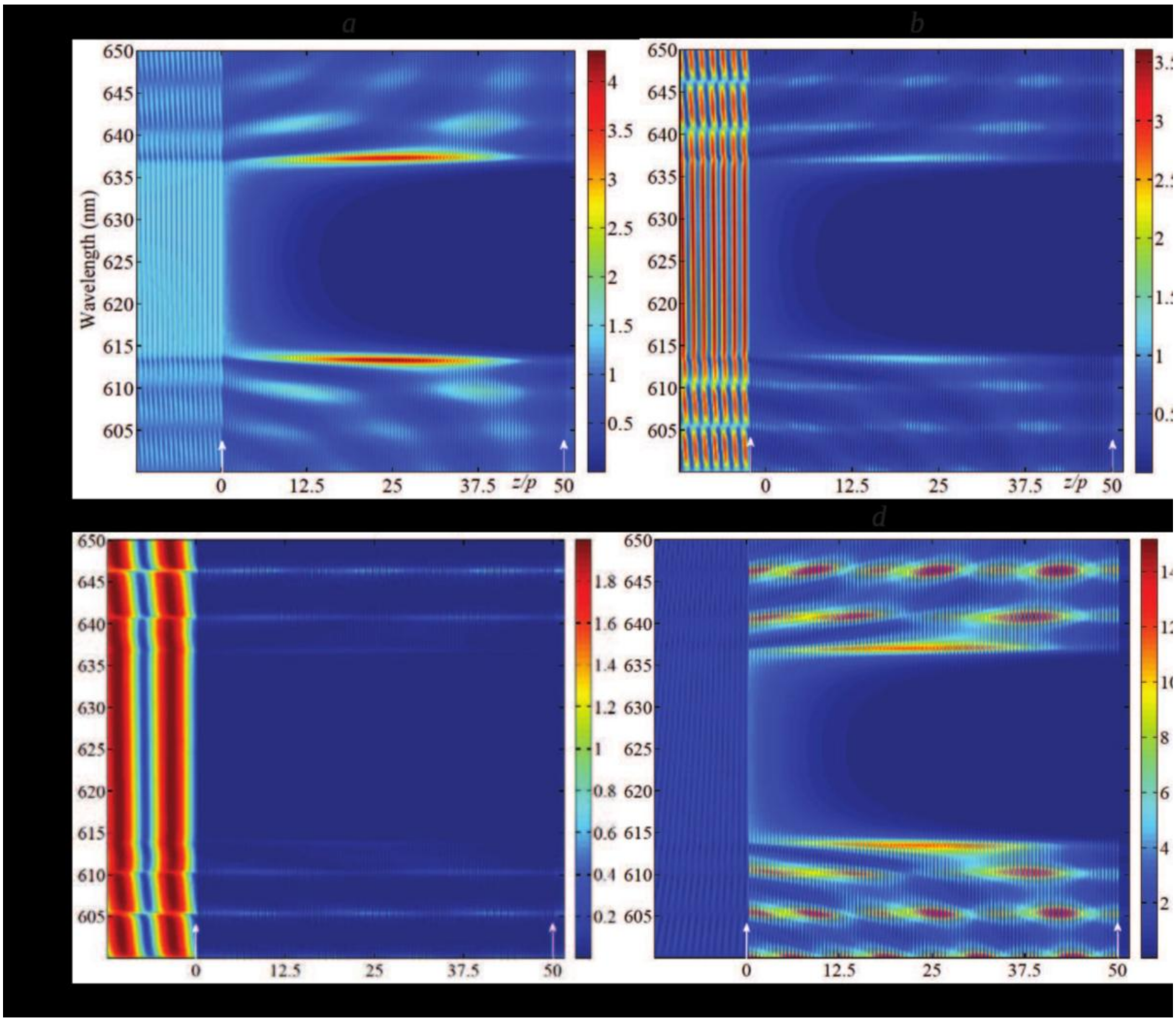

Fig. 5. (Color online). The density plot of the $\left|\vec{E}^{0,1,2}\right|^2$ spectra as a function of *z/p* for different values of $n_s = \sqrt{\varepsilon_s}$. *a*: $n_s = 1$; *b*: $n_s = 0.5$; *c*: $n_s = 0.1$; *d*: $n_s = 5$. *d* = 50*p*. The light incident on the CLC layer has a diffracting EP.

In Fig. 5, the evolution dependences of the spectra of $\left|\vec{E}^{0,1,2}\right|^2$ on the reduced coordinate, *z/p*, are presented for different values of $n_s = \sqrt{\varepsilon_s}$ :(*a*): $n_s = 1$; (b): $n_s = 0.5$; (c): $n_s = 0.1$; (*d*): $n_s = 5$.

The CLC layer occupies the range, $0 \le z/p \le 50$, (the CLC borders are denoted by the white arrows in the figure). The light incident on the CLC layer has a diffracting EP.

As it is seen from Fig. 5 (and also from Fig. 3), the total wave field for the diffracting EP in the PBG is evanescent, while outside the PBG $\left|\vec{E}^1\right|^2$ of the total wave in the CLC layer oscillates if *z* changes, and the beats are aroused. The beat minima coincide with the *z*=0 and *z*=*d* planes, at the minima of the reflection; and at the maxima of the reflection, the beat minima coincide with the plane, *z=0,* while the beat maxima coincide with the plane, *z* = *d*. There rises only one crest with a significant height at the first minimum of reflection. Here the total wave amplitude near the centre of the layer is much more than the amplitude of the incident wave amplitude. Due to the multiple reflections due to light diffraction in the finite volume an energy accumulation takes place at the layer centre. Two beat crests are aroused at the second reflection minimum, but they are of comparably lower height, etc.

One can also see from the figures in Fig. 5 that the maximum light energy accumulation in the layer takes place at the reflection minima of higher orders, if the difference, $\left|n_s - \sqrt{\varepsilon_m}\right|$, increases. It is seen from Fig. 5*c* that $\left|\vec{E}^1\right|^2 < \left|\vec{E}^0\right|^2$ at the reflection minimum, but the condition, $w^1 > w^0$, does take place here too, and in this case (i.e. in the case if $n_s << \sqrt{\varepsilon_m}$) low threshold lasing is possible at the reflection minima. On the contrary, it is seen from Fig. 5*d* that $\left|\vec{E}^1\right|^2 > \left|\vec{E}^0\right|^2$ at the reflection minimum, but here holds the condition, $w^1 < w^0$, and there is no low threshold lasing here.

In Fig. 6, the spectra of: $w^1 / w^0$; $\rho_2 / \rho_{iso}$ and $|A|$ – at absorption and gain – are presented. Then, the spectra of the ellipticity, *e*, and the azimuth, $\psi$, of the total wave in the CLC layer (at the border, *z* = 0) are brought in the same figure. As it is seen from this figure, the spectra of $w^1 / w^0$ and $\rho_2 / \rho_{iso}$ are very similar.

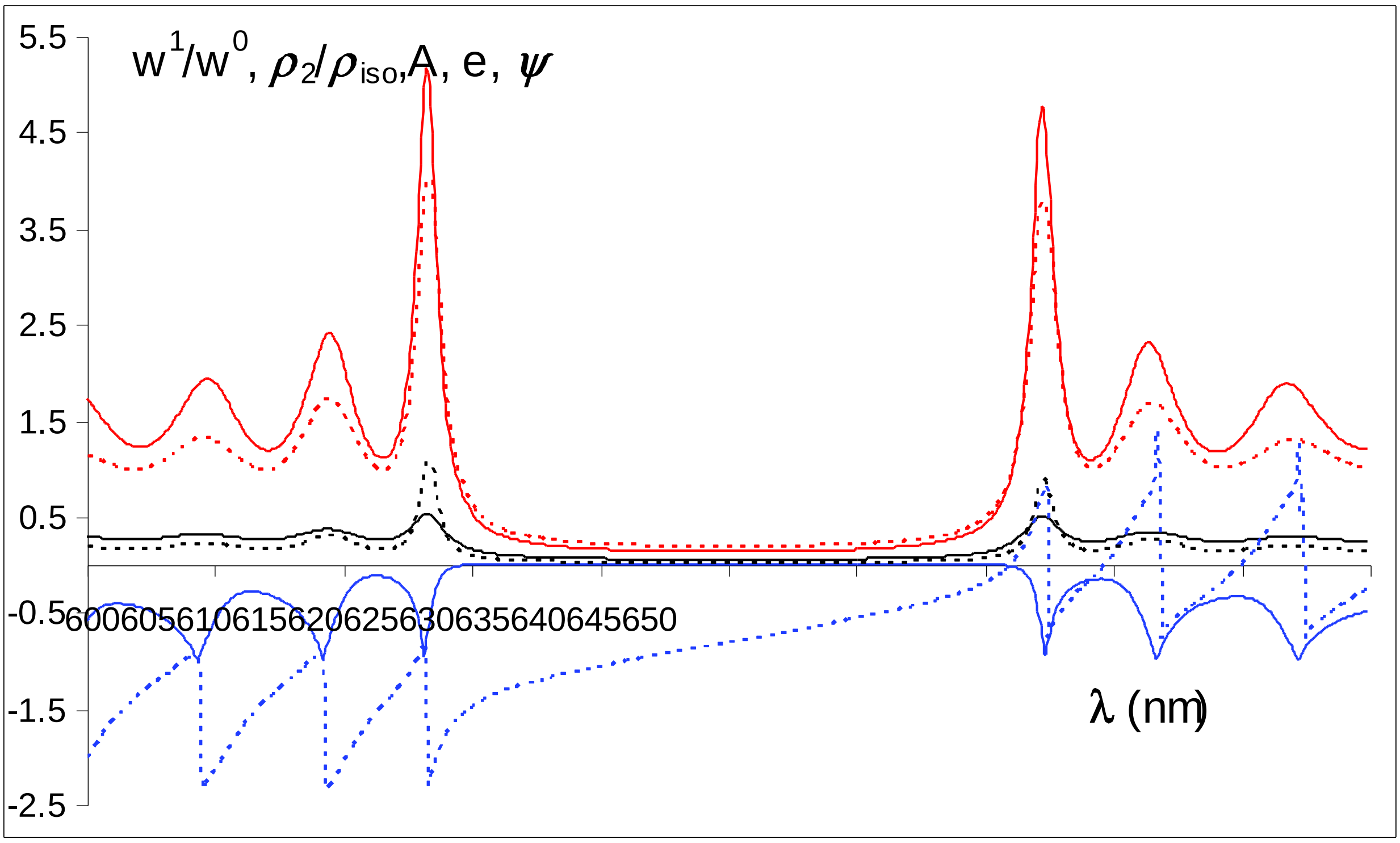

Fig. 6. The spectra of: $w^1 / w^0$ (the red dashed line); $\rho_2 / \rho_{iso}$ (the red solid line); *e* (the ellipticity, the blue solid line) and $\psi$ (the azimuth, the blue dashed line) of the total wave in the CLC (at the boundary, $z$ = 0), $A$ at absorption (the black solid line; $\mathrm{Im}\,\varepsilon_m = 0.002$ ) and $|A|$ at gain (the black dashed line; $\mathrm{Im}\,\varepsilon_m = -0.001$ ). The light incident on the CLC layer has a diffracting EP. $n_s = 1$ . $d$ = 50$p$.

At the low threshold lasing wavelengths, both the divergence in the PDS at the PBG edges of the CLC and that of the density of light energy accumulated in CLC layer take place, i.e. we can assert that: $w^1 / w^0 \propto \rho_2 / \rho_{iso}$ . This also is confirmed by the results presented below.

The density plots of the $|\vec{E}^{0,1,2}|^2$ spectra as a function of $z/p$ at absorption and gain are presented in Fig. 7.

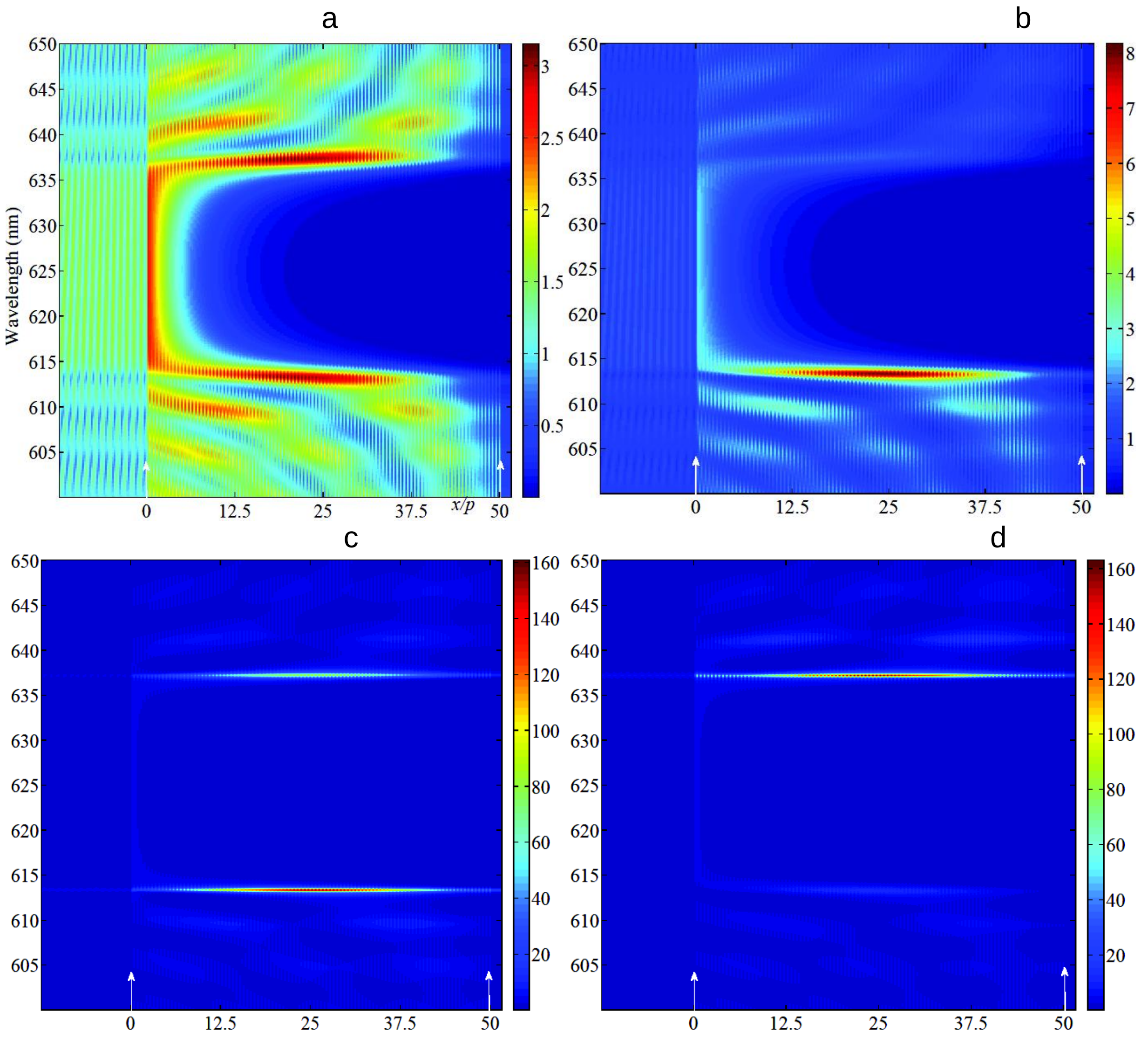

Fig. 7. (Color online). The density plot of the $\left|\vec{E}^{0,1,2}\right|^2$ spectra as a function of *z/p* at absorption and gain. *a*: $\mathrm{Im}\,\varepsilon_1 = \mathrm{Im}\,\varepsilon_2 = 0.003$; *b*: $\mathrm{Im}\,\varepsilon_1 = 0.006$, $\mathrm{Im}\,\varepsilon_2 = 0$; *c*: $\mathrm{Im}\,\varepsilon_1 = \mathrm{Im}\,\varepsilon_2 = -0.003$; *d*: $\mathrm{Im}\,\varepsilon_1 = -0.006$, $\mathrm{Im}\,\varepsilon_2 = 0$. $\varepsilon_s = \varepsilon_m$.

In Fig. 8, the evolutions of the spectra of $w^1$ if the gain coefficient, $y = \mathrm{Im}\,\varepsilon_m$, increases, are presented. (Compare the results presented in this figure with those in Fig. 4).

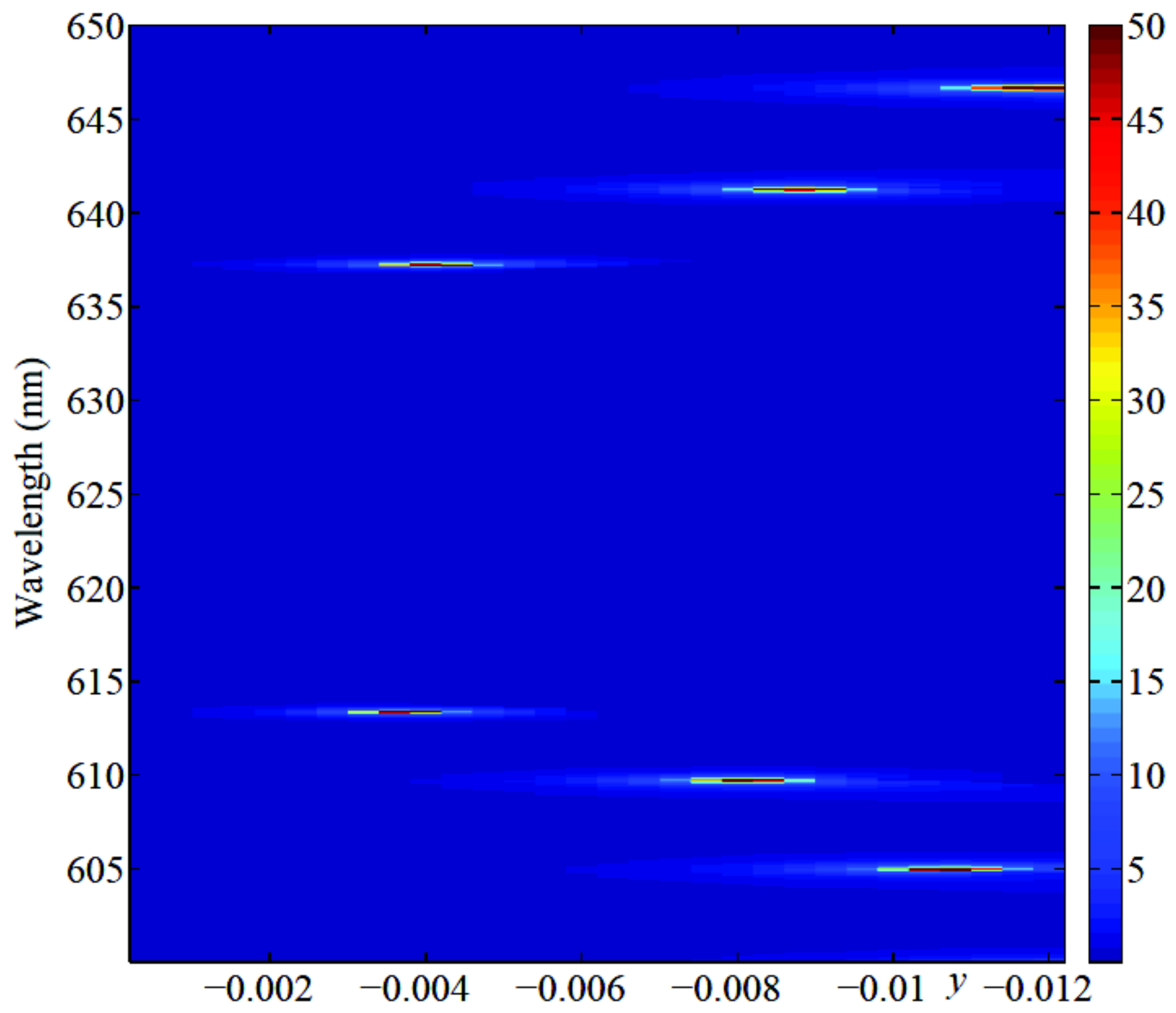

Fig. 8. (Color online). The density plot of the $w^1$ spectra as a function of the parameter, $y = \mathrm{Im}\,\varepsilon_m$, characterizing the gain. The incident light has a diffracting EP. $n_s = 1$. $d = 50p$.

Finally, we investigate the influence of the CLC layer thickness and its local anisotropy on the PDS of the diffracting EP. In Fig. 9, the dependence of the maximum PDS ($\rho_{2\max} / \rho_{iso}$) for the diffracting EP at the short wavelength edge on the reduced CLC layer thickness (on *d/p*) is presented.

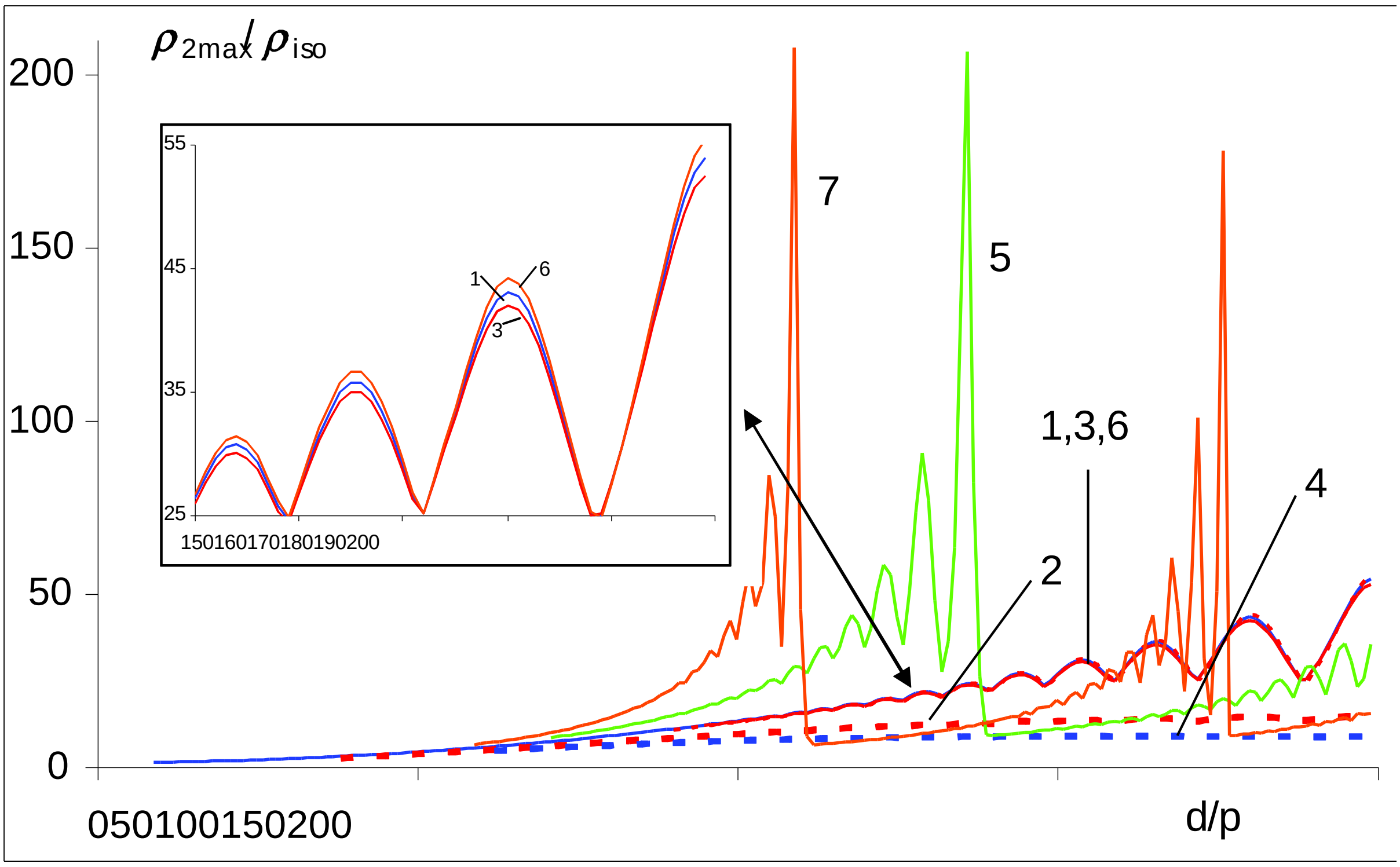

Fig. 9. (Color online).The dependence of the maximum PDS ( $\rho_{2\max} / \rho_{iso}$ ) for the diffracting EP at the SWE on the CLC layer reduced thickness (on *d/p*). 1. $\mathrm{Im}\,\varepsilon_1 = \mathrm{Im}\,\varepsilon_2 = 0$; 2. $\mathrm{Im}\,\varepsilon_1 = \mathrm{Im}\,\varepsilon_2 = 0.0002$; 3. $\mathrm{Im}\,\varepsilon_1 = 0.004,\ \mathrm{Im}\,\varepsilon_2 = 0$; 4. $\mathrm{Im}\,\varepsilon_1 = 0.,\ \mathrm{Im}\,\varepsilon_2 = 0.0004$; 5. $\mathrm{Im}\,\varepsilon_1 = \mathrm{Im}\,\varepsilon_2 = -0.0002$; 6. $\mathrm{Im}\,\varepsilon_1 = -0.004,\ \mathrm{Im}\,\varepsilon_2 = 0$; 7. $\mathrm{Im}\,\varepsilon_1 = 0.,\ \mathrm{Im}\,\varepsilon_2 = -0.0004$. $\varepsilon_s = \varepsilon_m$.

We discuss both the cases of the absence and presence of absorption and gain. Also, both the cases of isotropic and anisotropic absorption and gain are discussed. For the low values of the number of pitches, $N = d / p$, the dependence, $\rho_{2\max} / \rho_{iso}$, on $N$ is linear, but then this dependence becomes parabolic, and for the larger $N$ oscillations appear. Absorption leads to damping of these oscillations. When gain is present, the amplitudes of the oscillations monotonously increase. From a certain value of $N$ a sharp increment of the PDS takes place, and then the process is repeated. If an imaginary term is included only in the dielectric constant (which is parallel to the local director) then both absorption and gain practically do not have influence on the dependence of the PDS on $N$ at the short wavelength edge (naturally, for low absorption and gain). To complete this account we present in Fig. 10 the evolution of the PDS spectra when $N = d / p$ increase. As it is seen from Fig. 10, the PDS is maximum for the modes defined by condition (11). If the number of the mode, *m*, increases the maximum PDS decreases. Then, for every *m*, if the CLC thickness increases, the PDS increases too, but not monotonously, namely, with oscillations.

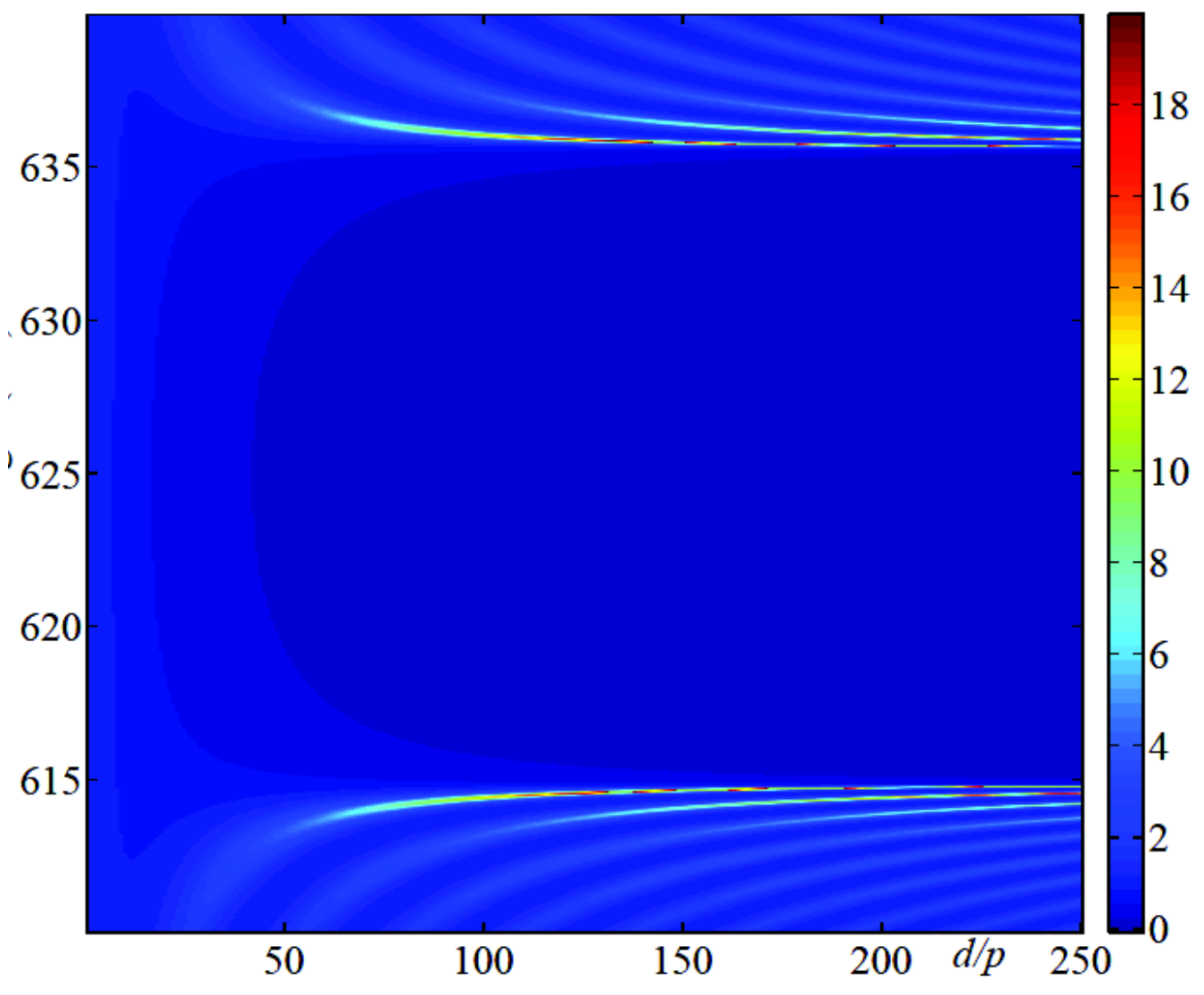

Fig. 10. (Color online). The density plot of the PDS spectra as a function of the parameter *d/p*. The incident light has a diffracting EP. $\varepsilon_s = \varepsilon_m$ .

In Fig. 11, the evolution of the PDS spectra, when $N = d / p$ increases, is presented at the presence of absorption (*a*) and gain (*b*) in the spectral range near the short wavelength border of PBG. The absorption, as it could be expected, leads to a decrement of the PDS, for every mode, if *N* increases, a PDS increment is possible too. There is an interesting situation at gain. As it is seen from this figure, for each mode, if *N* increases the PDS increases with oscillations, but there exists a critical value of *N* beyond which the lasing mode is quenched and the feedback vanishes. If the mode number, *m*, increases, then this critical value of *N* increases too.

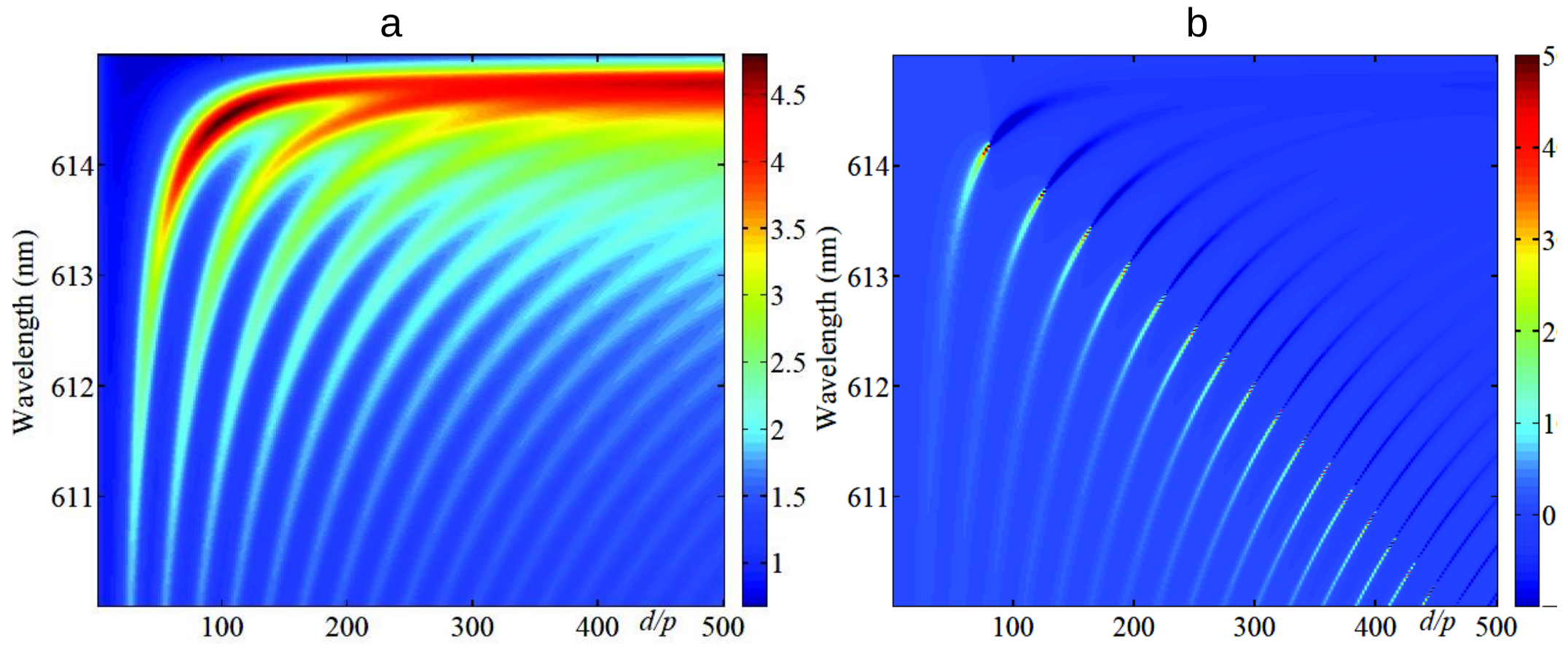

Fig. 11. (Color online). The density plot of the PDS spectra as a function of the parameter *d/p* at the presence of absorption (*a*: $\mathrm{Im}\,\varepsilon_m = 0.001$) and gain (*b*: $\mathrm{Im}\,\varepsilon_m = -0.001$). The incident light has a diffracting EP. $\varepsilon_s = \varepsilon_m$.

In Fig. 12, the dependence of the maximum PDS ($\rho_{2\max} / \rho_{iso}$) for the diffracting EP (at the short wavelength edge) on the CLC layer dielectric anisotropy, $\Delta = \dfrac{\varepsilon_1 - \varepsilon_2}{2}$, is presented. The blue line corresponds to the case, *d*=50*p*, and the red one is for the case, *d*=100*p*. As it is seen from the figure, when the anisotropy, *Δ* increases, the PDS monotonously increases, but then some irregular oscillations begin.

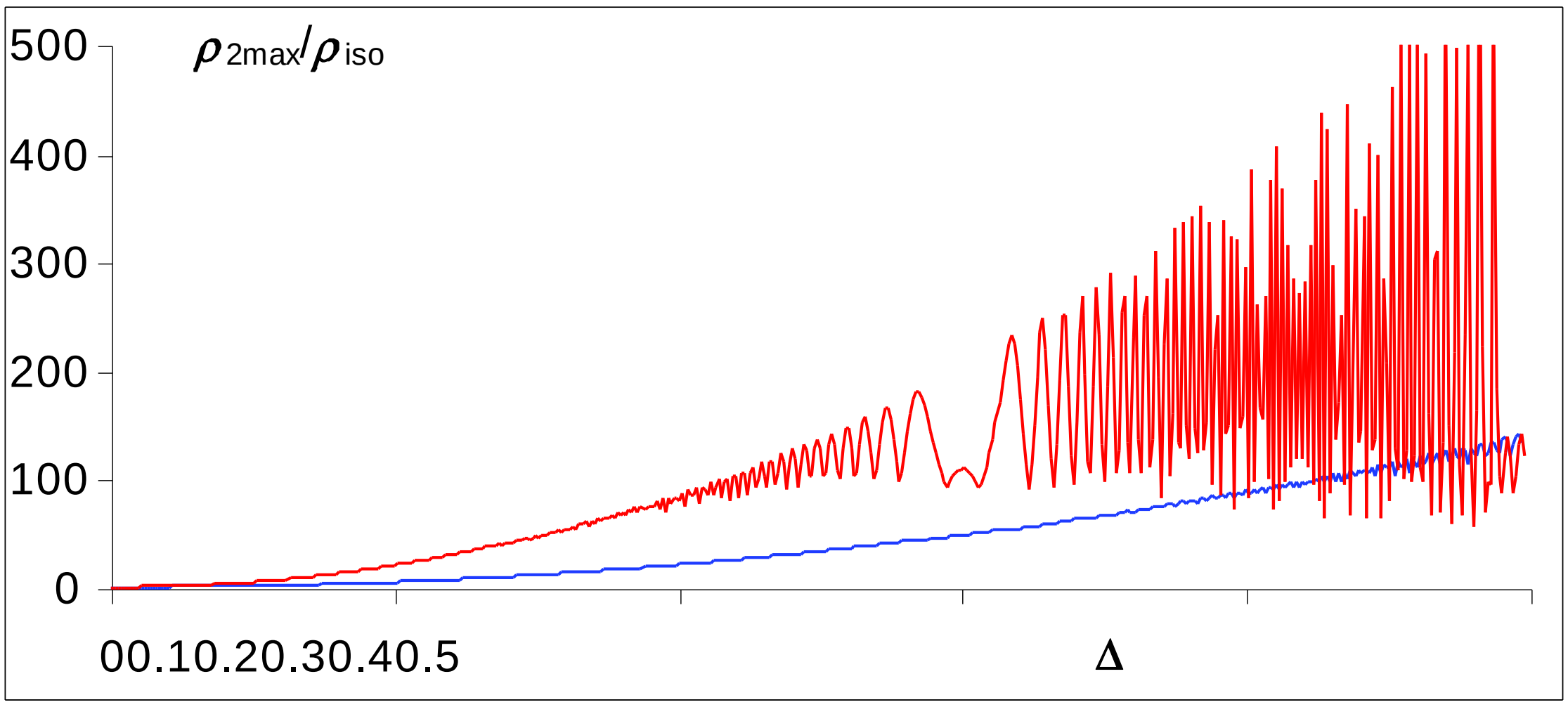

Fig. 12. (Color online). The dependence of the maximum PDS ($\rho_{2\max} / \rho_{iso}$) for the diffracting EP at the short wavelength edge on CLC dielectric anisotropy $\Delta = \dfrac{\varepsilon_1 - \varepsilon_2}{2}$ at the absence of absorption and gain. $n_s = 1$. $d = 50p$.

And, finally, in Fig. 13, we presented the evolution of the PDS spectra when the anisotropy $\Delta$ increases.

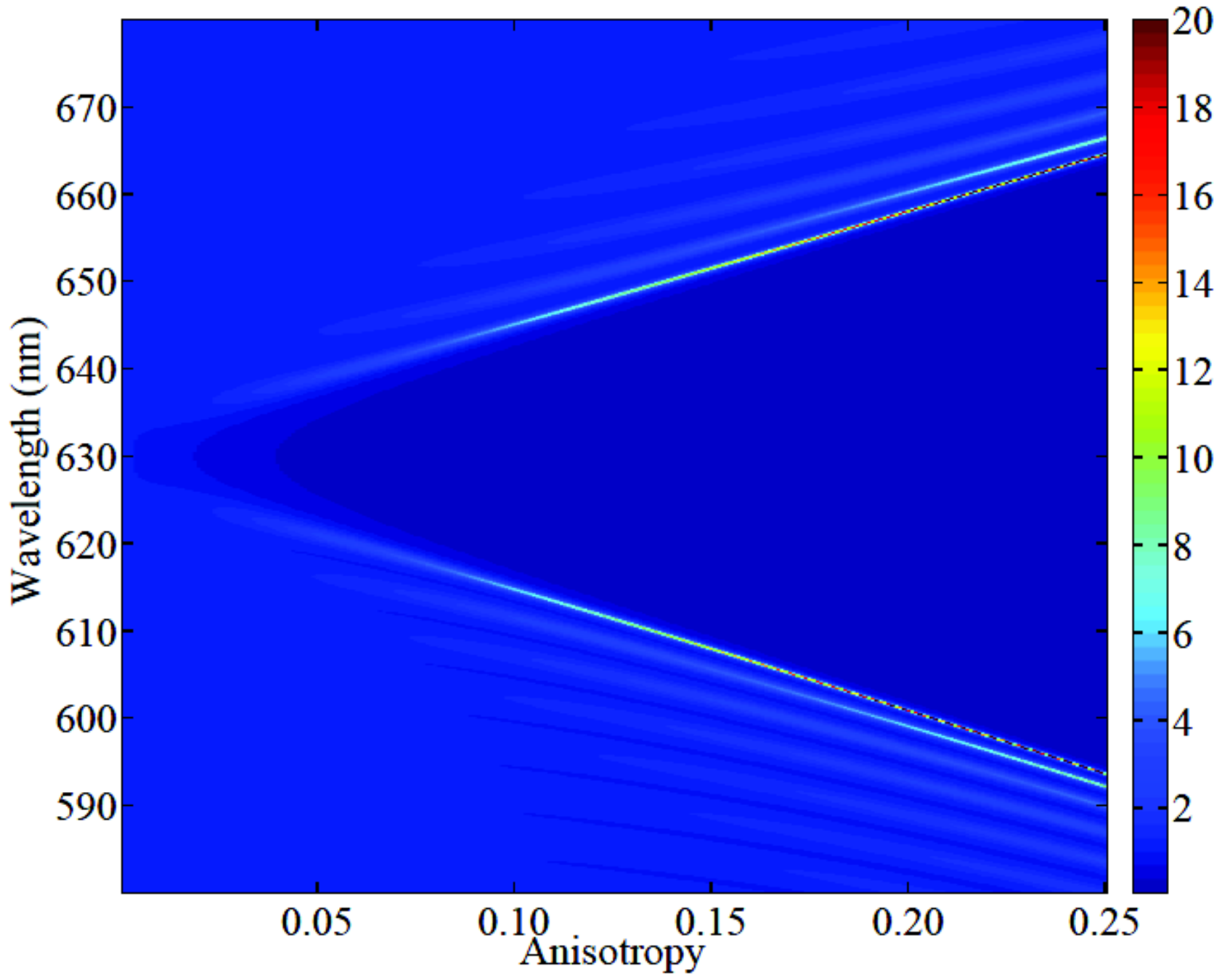

Fig. 13. (Color online). The density plot of the PDS spectra as a function of the anisotropy parameter at the absence of absorption and gain. The incident light has a diffracting EP. $\varepsilon_1 = 2.25 + \Delta$, $\varepsilon_2 = 2.25 + \Delta$. $\varepsilon_s = \varepsilon_m$. $d = 50p$.

## 4. Conclusions

We investigated the PDS peculiarities of the EPs for a CLC layer at absorption and gain, as well as the influence of: the CLC layer thickness; its dielectric anisotropy on the PDS. We obtained analytic expressions for the PDS of the EPs for a CLC layer of a finite thickness in the case of $\varepsilon_s = \varepsilon_m$ and $d = Np$, i.e. for the case of the minimum influence of the dielectric borders and if the layer thickness is a multiple of the helix pitch. These expressions are very complicated in the general case. We found out a number of new peculiarities, giving them plausible physical explanations. The influence of the gain on the PDS is investigated and it is shown that when the gain increases, the maximum PDS wavelengths move away from the PBG borders, and this takes place not continuously but in discrete paces. Then, we showed that there exist critical values of $x'$ beyond witch the lasing mode is quenching and the feedback vanishes. These critical values of $x'$ are different for SWE and LWE. The investigation of the influence of absorption on the PDS showed that the absorption leads to a decrement of the maximum PDS, which is natural. An interesting situation arouses when the absorption and gain are anisotropic. In particular, for the absorption, if an imaginary term is included only in the dielectric constant (which is parallel to the local director, in $\varepsilon_1$) then, if absorption increases, the PDS at SWE practically does not change at first, but afterwards it sharply decreases and if the the absorption continues to increase, the PDS sharply increases, too. It is shown that at the low threshold lasing wavelengths both a divergence in the PDS at the band edges of the CLC layer and a divergence of density of light energy accumulated in it take place. We showed that the dependences of the PDS on the effective layer thickness oscillate for the larger values of the latter. We showed that at the presence of gain there exists a critical value of numbers of pitches in the CLC layer beyond which the lasing mode is quenched and the feedback vanishes, too.